\documentclass[structabstract]{aa}
\usepackage{graphicx}  
\usepackage{txfonts}
\usepackage{subfigure}
\usepackage{natbib}
\bibpunct{(}{)}{;}{a}{}{,} 

\def\ll_lsun{log$({L/\rm L_{\odot}})$~}  
  
\def\masa_msun{$M/ \rm M_{\odot}$~}  
  
\def\m_mstar{$M/M_{*}$~}

\def\v4334{\mbox{V4334 Sgr}}

\def\112{\mbox{SDSS J111215.82+111745.0}}  
\def\840{\mbox{SDSS J184037.78+642312.3}}  
\def\518{\mbox{SDSS J151826.68+065813.2}}  
\def\614{\mbox{SDSS J161431.28+191219.4}}  
\def\228{\mbox{SDSS J222859.93+362359.6}}
\def\738{\mbox{PSR J1738+0333}}
\def\J1618{\mbox{SDSS~J161831.69+385415.15}}

\begin{document}  

\title{The coolest extremely low-mass white dwarfs}

\author{Leila M. Calcaferro\inst{1,2},
  Leandro G. Althaus\inst{1,2}, \and
        Alejandro H. C\'orsico\inst{1,2}}  
\offprints{L. M. Calcaferro} 

\institute{$^1$ Grupo  de Evoluci\'on  Estelar y  Pulsaciones,  Facultad de 
           Ciencias Astron\'omicas  y Geof\'{\i}sicas, Universidad
           Nacional de La Plata, Paseo del Bosque s/n, (1900) La
           Plata, Argentina\\   $^{2}$ Instituto de Astrof\'{\i}sica
           La Plata, CONICET-UNLP, Paseo  del Bosque s/n, (1900) La
           Plata,
           Argentina\\  \email{lcalcaferro,althaus,acorsico@fcaglp.unlp.edu.ar}}
\date{Received}  

\abstract{Extremely low-mass white dwarf (ELM WD; $M_{\star} \lesssim 0.18-0.20 M_{\sun}$) stars are thought to
  be formed in binary systems via  stable or unstable mass transfer. 
 Although stable mass transfer predicts the formation of ELM WDs
with thick hydrogen (H) envelopes, and hence characterized by dominant residual nuclear burning along
the cooling branch,
the formation of ELM WDs with thinner H envelopes from unstable
mass loss cannot be discarded.}
{ We compute new evolutionary sequences for helium (He) core WD stars with thin H envelope with the main aim of assessing 
  the lowest $T_{\rm eff}$ that could be reached by this type of stars.}
  { We generate a new grid of evolutionary sequences of He core WD stars
  with thin H envelope in the mass range from $0.1554$ to
  $0.2025 M_{\sun}$, and assess the changes in both the cooling times and surface gravity induced by a reduction of the
  H envelope. We also determine, taking into account the predictions of progenitor evolution, the lowest $T_{\rm eff}$
  reached by the resulting ELM WDs.}
  { We find that a slight reduction in the
  H envelope yields a significant increase in the cooling rate of ELM WDs. 
  Because of this,  ELM WDs with thin H envelope
  could cool down to $\sim 2500\ $K, in contrast with their canonical
  counterparts that cool down to $\sim 7000\ $K. In addition, we find that a reduction
  of the thickness of
  the H envelope increases markedly the surface gravity (g) of these stars.}
  {If ELM
  WDs are formed with thin H envelopes, they could be detected at very
  low $T_{\rm eff}$.  The detection of such cool
  ELM WDs would be indicative that they were formed with
  thin H envelopes, thus opening the possibility of placing constraints
  to the possible mechanisms of formation of this type of stars. Last but
  not least, the
  increase in $g$ due to the reduction of the H envelope leads to
  consequences in the spectroscopic determinations of these stars. }
   \titlerunning{Asteroseismology of ELMV WDs}   \maketitle
\authorrunning{Calcaferro et al.}  

   
\section{Introduction}  
\label{intro}  

White dwarf  (WD) stars constitute  the  most common fate in stellar evolution. Indeed,
more than 90\% of all stars born in the galaxy will ultimately evolve into WDs,
earth-sized, electron degenerate objects. 
These stars play a central and unique role in our understanding of the formation and
evolution of stars, our galaxy itself, and planetary systems. In addition, they constitute
valuable laboratories of extreme physics. Their collective properties
allow us to infer valuable information  about  the star formation history of
the  Solar neighborhood,  AGB mass loss, and   to  study  the properties  of
various   stellar   populations  --   see   \cite{2008PASP..120.1043F,
2008ARA&A..46..157W}, and  \cite{2010A&ARv..18..471A} for  reviews. In
particular, WDs are  used as accurate  age indicators  for a
wide variety of Galactic populations, including the disk, and open and
globular      clusters      --     see      \cite{2009ApJ...693L...6W,
2010Natur.465..194G,     2011ApJ...730...35J,     2013A&A...549A.102B,
2013Natur.500...51H}  and \cite{2015A&A...581A..90T}  for some  recent
applications.

The WD mass distribution comprises  a  population of low-mass
remnants, most of them expected to have a He  core.
These stars are thought to be the result of strong  mass-loss episodes in interactive
binary systems before the He flash during the red giant branch phase
of low-mass  stars, see \cite{2000MNRAS.316...84S,2013A&A...557A..19A,2016A&A...595A..35I}, and references therein.
Specifically, this interactive binary evolutionary scenario is thought to be the  most
plausible origin for  the so-called  extremely low-mass (ELM) WDs,
which  have masses below $\sim  0.18-0.20 M_{\sun}$\footnote{In this paper
  we will refer ELM WDs to those He-core WDs  for which no CNO H shell flashes
  are expected during the cooling phase.}.

In recent years, the number of observed low-mass WDs, including ELM WDs, 
has  increased considerably due to the result of many ELM surveys, and the  SPY and  WASP surveys
\citep[see][]{2009A&A...505..441K,  2010ApJ...723.1072B,
  2012ApJ...744..142B, 2011MNRAS.418.1156M, 2011ApJ...727....3K,
  2012ApJ...751..141K, 2013ApJ...769...66B, 2014ApJ...794...35G,
  2015MNRAS.446L..26K, 2015ApJ...812..167G}.
 The evolution of  He-core WDs  resulting from binary evolution has recently been studied in detail  by  \cite{2013A&A...557A..19A} and
 \cite{2016A&A...595A..35I}.
 In addition, the detection of
  pulsation $g$ modes (gravity modes) in some of these stars   \citep{2012ApJ...750L..28H,
  2013ApJ...765..102H,2013MNRAS.436.3573H,
  2015MNRAS.446L..26K,2015ASPC..493..217B,2017ApJ...835..180B} has
  given rise to a new class of variable WDs, the ELMVs.
  These pulsating low-mass WDs provide us an unique chance for
  probing the interiors of  these stars and possibly to test
  their   formation scenarios by employing the tools of
  asteroseismology.

  The  accepted mechanism  for  the  formation of low-mass He-core
  WDs is  either
  through unstable mass loss via common-envelope episodes or stable mass
  loss via Roche-lobe
  overflow   in  close   binary  systems   \citep[see  for   a  recent
    discussion][]{2016A&A...595A..35I}. In particular, existing evolutionary
  tracks for
  ELM  WDs are  derived from  progenitor stars  that have  experienced
  stable               mass                transfer,               see
  \cite{2013A&A...557A..19A,2016A&A...595A..35I}     and    references
  therein.  All  of these   studies  predict for these stars the occurrence
  of  thick
  H envelopes  that sustain  residual stable  H burning,
  thus yielding  extremely long  cooling ages  even at  high effective
  temperatures. In particular, these studies show that no ELM (those which
  do not experienced CNO H flashes) is
  expected to have cooled below $\sim 7000 K$.

  However, the  formation of low-mass  He-core WDs with  thin H
  envelopes   unable  to   sustain  residual   H  burning   cannot  be
  discarded. Such  WDs could result from  common-envelope evolution of
  close               binary                systems,               see
  \cite{2016MNRAS.460.3992N,2016MNRAS.462..362I,2017MNRAS.470.1788C} for recent
  calculations,  or from the lost of the envelope of a RGB star induced by an
  inspiralling                       giant                      planet
  \citep{1998A&A...335L..85N,2002PASP..114..602D,2017arXiv170608897S}.
  These studies suggest that  most of  the
  envelope of a RGB star could indeed be lost in these episodes. In
  particular, \cite{2017MNRAS.470.1788C} found that  the envelope
  of their low-mass  RGB models becomes dynamically  unstable, with the
  result that  the entire  envelope of  the star  is removed  over the
  duration  of  the  slow  spiral-in phase.   However,  it  should  be
  mentioned that the prediction of  such channels for the formation of
  ELM WDs  is more uncertain, because  of the large binding  energy of
  the donor's  envelope in  this case  \citep{2017arXiv170301648S}. On
  the observational  side, the  existence of  a population  of low-mass
  He-core  WDs with  thin H  envelopes in  NGC 6397  is not
  discarded, see  \cite{2009ApJ...699...40S}. Finally, in  an entirely
  different  context, the  formation  of ELM  WDs  with thin  H
  envelope could result from the  irradiation of the pulsar companion,
  leading      to     a      fast     cooling      of     the      ELM
  \citep{1988Natur.334..227V,2001MNRAS.321...71E}.

 In view of these considerations, we cannot entirely rule out the
 existence of ELM WDs with thin H envelopes.  To explore the
 impact of this possible outcome of binary interaction on  observational expectations of ELM WDs ,
 we present in this paper new evolutionary sequences for ELM WDs
 formed with thin H envelope. We will show that in this case
 the cooling times of such WDs turn out to be much shorter than the
 cooling times of their counterparts with thick (canonical) H
 envelopes, that predict that no ELM WD is expected to be observed
 below $\sim 7000 K$, in agreement with the current observational
 status.  Thus, according to our calculations, if ELM WDs were born
 with thin  H envelope (for instance, due to unstable mass
 transfer) they could be observed at much lower effective temperatures
 (down to $2500\ $K).  The detection of ELM WDs  at low effective
 temperatures  would be indicative that such WDs were formed with thin
 H envelope,  a fact that  could shed light on the nature of
 mass loss that leads to the formation of  ELM WDs.  Although the
 detection of such cool ELM WDs is difficult, because the spectroscopic
 technique becomes inaccurate below $7000\ $K where the Balmer lines
 become weaker and then disappear below $5000\ $K, parallax measurements
 from Gaia is a promising avenue that would help to identify such cool
 ELM WDs.

This paper is organized as follows. A brief description  of the
stellar models and the numerical code employed is provided in
Sect. \ref{evolutionary}.  In Sect. \ref{analysis} we study the cooling
times for some selected ELM WD sequences with thin and thick H envelope.
Next, we describe the considerations we take into account to find the
lowest effective temperature that could be reached by ELM WD stars
considering different possible progenitors and we show the results
obtained.
Finally, in Sect. \ref{conclusions} we summarize the main
findings of this work.

\section{Evolutionary models}  
\label{evolutionary}  
This work is based on fully evolutionary models of
low-mass He-core WDs generated with  the {\tt LPCODE} stellar
evolution code.  This code computes in detail the complete
evolutionary stages that lead to the WD formation, allowing the study
of the WD evolution consistently  with the predictions of the
evolutionary  history of progenitors.  Details of the {\tt LPCODE} can be
found in \citet{2005A&A...435..631A,
2009A&A...502..207A,2013A&A...557A..19A,2015A&A...576A...9A} and
references therein. Here, we briefly mention  the   ingredients
employed  that  are  relevant for  our analysis  of  low-mass
He-core WDs \citep[see][for details]{2013A&A...557A..19A}.  The
standard  mixing length  theory (MLT)   for convection in the ML2
prescription is used
\citep[see][for its definition]{1990ApJS...72..335T}.
We assume the metallicity of the progenitor stars to be $Z = 0.01$.
 For the WD regime, we consider  the radiative opacities for arbitrary metallicity in
the range of 0 to  0.1 from the OPAL
project \citep{1996ApJ...464..943I}. Conductive opacities are those
of \citet{2007ApJ...661.1094C}. For the main-sequence evolution, we
consider the equation of state from OPAL for H- and He-rich
compositions.   Neutrino  emission   rates   for   pair,
 photo,   and bremsstrahlung processes are those of 
\citet{1996ApJS..102..411I}, and  for plasma processes, those
of   \citet{1994ApJ...425..222H}. For the WD
regime we have employed  an updated version of
the equation of state of \citet{1979A&A....72..134M}. The nuclear network
takes into account 16 elements and 34 thermonuclear  reaction  rates, most
of them corresponding to  pp-chains and  CNO bi-cycle,  needed
for this work. We also
consider time-dependent diffusion due to gravitational  settling and
chemical  and thermal diffusion of nuclear  species  following  the
multicomponent  gas  treatment  of
\citet{1969fecg.book.....B}. We have computed abundance changes
according to element diffusion, nuclear reactions,  and convective
mixing, a treatment  that represents a very significant aspect in
evaluating the importance of residual nuclear burning during the
cooling stage of low-mass WDs.

Realistic configurations for our initial low-mass He-core WD models  were 
taken from binary evolutionary calculations  \citep{2013A&A...557A..19A}. There, binary
evolution was assumed to be fully nonconservative, and the losses of
angular momentum due to mass loss, gravitational wave radiation, and
magnetic braking were considered. Binary configurations assumed in
\cite{2013A&A...557A..19A} consist of an evolving main-sequence low-mass component (donor
star) of initially $1 M_{\sun}$ and a $1.4 M_{\sun}$ neutron star
companion as the other component. Initial He-core WD
models with stellar masses ranging from $0.1554$ to $0.4352\ M_{\sun}$ characterized by thick H envelopes were derived from stable mass loss via Roche-lobe overflow, see \cite{2013A&A...557A..19A}
for details.
The evolution of these models was
computed down to the range of luminosities of cool WDs, including the
stages of multiple thermonuclear CNO flashes at the beginning of
the cooling branch.

Initial ELM WD models with thin H envelopes were generated
from those computed in \cite{2013A&A...557A..19A}, as mentioned above, by
artificially reducing the thickness
of the H envelope at high luminosities, during the pre-white dwarf evolutionary stages. This ensures that  all the transitory effects associated
with this artificial procedure have already ceased
by the time the resulting remnants reach the cooling branch. For the purpose of getting
different thicknesses of the H envelope, for each sequence characterized
by a given $M_{\star}$ and a thick (canonical) value of $M_{\rm H}$, as
predicted by the full computation of the pre-WD evolution, we simply
replaced $^1$H by $^4$He from a given mesh point to get the desired
H envelope mass. Ongoing element diffusion smoothes the chemical
profile at the H/He chemical transition region before the remnant 
settles onto the cooling
branch. With the aim of exploring
not only the impact of the H envelope on the cooling time,  for which only a small reduction 
in the envelope thickness is required, but also on the surface
gravity, we have generated additional sequences with H envelope several orders of
magnitude thinner than the canonical value.
We have also generated two evolutionary sequences with $0.130$ 
and $0.140\ M_{\sun}$ with thin H envelopes to cover the domain of low
surface gravities of our grid. This was done by
artificially scaling the total mass of the $0.1554\ M_{\sun}$
white dwarf model.

\section{Searching for ELMs with the lowest effective temperature}  
\label{analysis}

\begin{figure}  
\centering   \includegraphics[clip,width=230pt]{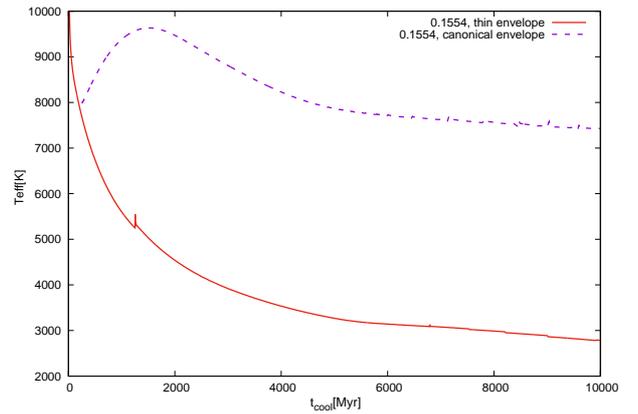}  
\caption{Cooling curves for the ELM WDs with $M_{\star}=   0.1554\ M_{\sun}$
  with both canonical and thin H envelopes.}
\label{fig:tcool_teff_0.1554}
\end{figure}

\begin{figure}  
\centering   \includegraphics[clip,width=230pt]{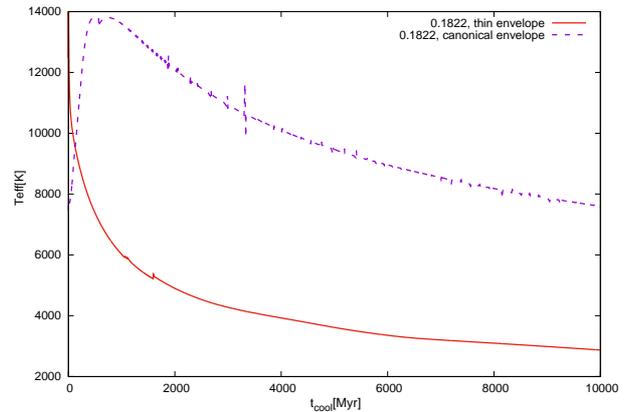}  
\caption{Same as Fig.~\ref{fig:tcool_teff_0.1554} but for the evolutionary
  sequences with  $M_{\star}=   0.1822\ M_{\sun}$.}
\label{fig:tcool_teff_0.1822}
\end{figure}

 We begin by examining the impact of the H envelope mass on the
cooling times of the ELM WDs. In this sense, we find that a small 
reduction in the H envelope, by a factor of 2 - 3, is enough to turn off
the H burning,  thus leading to a fast evolution at advanced stages
\citep[see][for a similar result]{2001MNRAS.323..471A}. This is illustrated
in Figs.~\ref{fig:tcool_teff_0.1554} and
\ref{fig:tcool_teff_0.1822}, which display the cooling times as a function
of the effective temperature for the sequences with $0.1554$ and $0.1822\ M_{\sun}$,
respectively. We stress that these sequences did not experience CNO flashes on the
cooling branch. Cooling times are shown for both canonical and
thin H envelopes. Note from the figures the fast cooling that characterize 
sequences with thin H envelopes, as compared with their counterparts with canonical 
envelopes powered by residual H burning. In particular, for the $M_{\star}= 0.1554\ M_{\sun}$
canonical sequence, evolution takes roughly $8700\ $Myr to cool down 
to an effective temperature of $7500\ $K, in contrast with the $290\ $Myr required by the
thin envelope sequence. More specific information is provided
by Table~\ref{tab:cooling}, which tabulates the cooling time,
$t_{cool}$, along with $\log(g)$ for some selected  $0.1554$, $0.1822$,
and $0.2025\ M_{\sun}$
models having canonical (with $M_{\rm H}= 2.0 \times 10^{-2}$,
$3.6 \times 10^{-3}$ and $3.7 \times 10^{-3}\ M_{\star}$ for the $0.1554$,
$0.1822$, and $0.2025\ M_{\sun}$ sequences,  respectively) and thin
H envelopes
($M_{\rm H} \sim 10^{-5}\ M_{\star}$ for all the sequences).
We mention that the $0.2025\ M_{\sun}$ 
sequence experiences CNO-flashes during the first cooling branch, with the consequent
result that the H envelope and H burning on the final
cooling branch are reduced. This explains the short cooling age of this sequence
also in  the case of canonical envelope.
It is clear that residual nuclear burning
is responsible for the extremely large cooling times that would need ELM WDs
to reach low effective temperatures, and more importantly,  that a small reduction of
the H envelope is enough to extinguish this nuclear burning source,
and thus cause a fast evolution of the ELM WD.

\begin{table*}[t]
\centering
\caption{Main characteristics ($T_{\rm eff}$, $\log(g)$ and $t_{\rm cool}$) for some
  selected $0.1554$, $0.1822$ and $0.2025\ M_{\sun}$ models having canonical and thin envelopes.}
\begin{tabular}{c|ccc|ccc}
\hline
\hline
 $M_{\star}[M_{\sun}]$& $T_{\rm eff}$[K] & $\log(g)$[cgs] & $t_{\rm cool}$ [Gyr] & $T_{\rm eff}$[K] & $\log(g)$[cgs] & $t_{\rm cool}$ [Gyr]\\
\hline
\noalign{\smallskip}
& &Thick envelope &  & & Thin envelope \\
\noalign{\smallskip}
\hline
$0.1554$ &$9420$ & $5.682$ & $2.077$ &$9440$ &$6.287$ & $0.032$\\
         &$8930$ & $5.858$ & $2.811$ &$8940$ &$6.358$ & $0.058$\\
         &$8420$ & $5.995$ & $3.633$ &$8430$ &$6.436$ & $0.115$\\
         &$7930$ & $6.108$ & $4.742$ &$7930$ &$6.505$ & $0.203$\\
         &$7430$ & $6.245$ & $10.12$ &$7430$ &$6.561$ & $0.309$\\
         &       &         &         &$6930$ &$6.611$ & $0.444$\\
         &       &         &         &$6430$ &$6.654$ & $0.608$\\
         &       &         &         &$5930$ &$6.694$ & $0.820$\\
         &       &         &         &$5430$ &$6.737$ & $1.112$\\
         &       &         &         &$4930$ &$6.812$ & $1.593$\\
         &       &         &         &$4430$ &$6.834$ & $2.145$\\
         &       &         &         &$3930$ &$6.855$ & $2.968$\\
         &       &         &         &$3430$ &$6.875$ & $4.356$\\
         &       &         &         &$2930$ &$6.912$ & $8.469$\\
         &       &         &         &$2540$ &$6.927$ & $12.21$\\
\hline
$0.1822$ &$10430$ & $6.479$  & $3.583$ & $10420$ & $6.548$ & $0.052$\\
         &$9930$  & $6.516$  & $4.249$ & $9930$  & $6.595$ & $0.087$\\
         &$9430$  & $6.554$  & $5.066$ & $9430$  & $6.642$ & $0.139$\\
         &$8930$  & $6.590$  & $6.074$ & $8940$  & $6.684$ & $0.203$\\
         &$8420$  & $6.624$  & $7.203$ & $8440$  & $6.722$ & $0.279$\\
         &$7940$  & $6.655$  & $8.848$ & $7930$  & $6.756$ & $0.374$\\
         &$7430$  & $6.685$  & $10.69$ & $7440$  & $6.787$ & $0.486$\\
         &$6930$  & $6.713$  & $13.03$ & $6920$  & $6.816$ & $0.631$\\
         &        &          &         & $6440$  & $6.844$ & $0.808$\\
         &        &          &         & $5930$  & $6.872$ & $1.046$\\
         &        &          &         & $5430$  & $6.904$ & $1.402$\\
         &        &          &         & $4930$  & $6.967$ & $1.971$\\
         &        &          &         & $4420$  & $6.986$ & $2.682$\\
         &        &          &         & $3930$  & $7.008$ & $4.004$\\
         &        &          &         & $3430$  & $7.021$ & $5.675$\\
         &        &          &         & $2930$  & $7.039$ & $9.497$\\
         &        &          &         & $2550$  & $7.046$ & $12.18$\\
\hline
$0.2025$ &$10430$  & $6.695$  &$0.003$ &$10430$ &$6.798$ & $0.004$\\
         &$9910$   & $6.706$  &$0.059$ &$9940$  &$6.809$ & $0.007$\\
         &$9420$   & $6.719$  &$0.121$ &$9440$  &$6.825$ & $0.014$\\
         &$8950$   & $6.735$  &$0.188$ &$8940$  &$6.844$ & $0.034$\\
         &$8430$   & $6.753$  &$0.278$ &$8450$  &$6.868$ & $0.093$\\
         &$7930$   & $6.771$  &$0.380$ &$7910$  &$6.895$ & $0.199$\\
         &$7430$   & $6.790$  &$0.499$ &$7450$  &$6.916$ & $0.314$\\
         &$6930$   & $6.811$  &$0.650$ &$6960$  &$6.937$ & $0.463$\\
         &$6440$   & $6.833$  &$0.849$ &$6430$  &$6.960$ & $0.673$\\
         &$5930$   & $6.859$  &$1.117$ &$5930$  &$6.981$ & $0.924$\\
         &$5430$   & $6.889$  &$1.480$ &$5430$  &$7.007$ & $1.284$\\
         &$4930$   & $6.941$  &$2.031$ &$4930$  &$7.058$ & $1.886$\\
         &$4430$   & $6.998$  &$2.844$ &$4420$  &$7.075$ & $2.729$\\
         &$3930$   & $7.031$  &$4.418$ &$3930$  &$7.093$ & $4.132$\\
         &$3430$   & $7.047$  &$6.166$ &$3430$  &$7.113$ & $6.810$\\
         &$2930$   & $7.058$  &$8.525$ &$2940$  &$7.120$ & $9.502$\\
         &$2530$   & $7.064$  &$11.17$ &$2530$  &$7.124$ & $12.13$\\
\hline
\end{tabular}
\label{tab:cooling}
\end{table*}

 The short cooling times that characterize the sequences with thin
H envelopes  opens
the possibility of detecting ELM WDs at very low $T_{\rm eff}$.
To place this on a quantitative basis, we will assume an age of the
Galactic disk of $13.7\ $Gyr.  If we take  $t_{\rm birth}$ and $t_{\rm MS}$ as the
time of birth and the main sequence time of the progenitor star, respectively, 
then  the available
time for the ELM WD to evolve during the WD stage is given by
$13.7 Gyr - t_{\rm MS} - t_{\rm birth}$.
 To obtain the maximum available time on the WD regime, and hence to find the 
lowest $T_{\rm eff}$ that may be reached by a low-mass He WD, 
we set $t_{\rm birth}= 0$, i.e. we assume that the progenitor was born at the
beginning of star formation in the disk. In addition,
%
the shortest ages for the possible progenitors of these stars must be considered. To assess this, we computed 
additional evolutionary calculations to estimate the maximum value of the initial mass of the progenitor stars at the ZAMS
that may form ELM WDs, and we have obtained a mass of $\sim 1.5\ M_{\sun}$ ( $t_{\rm MS} \sim 1.5$ Gyr for solar metallicity).
This is in line with the exploration made by \cite{2017arXiv170301648S}.
 Taking into account the $t_{\rm MS}$ of this progenitor, we explore the lowest  $T_{\rm eff}$ values
that can be reached by the  $M_{\star}= 0.1554\ M_{\sun}$ and
$M_{\star}= 0.1822\ M_{\sun}$  non-flashing sequences, and the $M_{\star}= 0.2025\ M_{\sun}$ CNO-flashing sequence.

\begin{figure*}  
\centering   \includegraphics[clip,width=420pt]{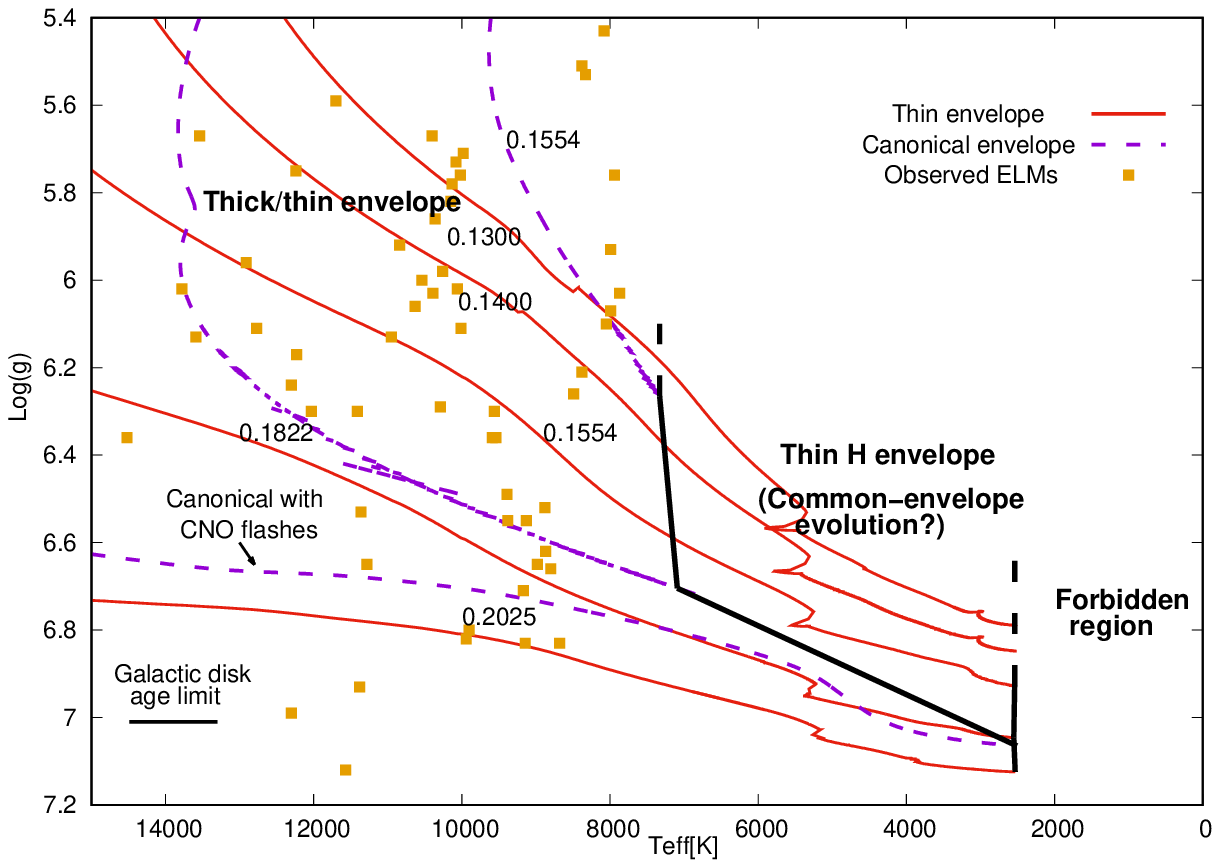}  
\caption{Evolutionary sequences of low-mass WDs with different
  thicknesses of the H envelope
  (canonical and thin) in the $\log(g)$ vs $T_{\rm eff}$ plane.  The vertical black
  line to the left (right) connects points of
  equal age, $13.7\ $Gyr, for the canonical (thin H envelopes) tracks. The dashed lines also correspond to the galactic
  disk age limit, and are only marked as projections to delimit the
  region. We have  also included the $0.130$ and  $0.140\ M_{\sun}$ artificial evolutionary sequences. Also included are the observed ELMs and ELMVs. The progenitor considered has $M_{\star}= 1.5\ M_{\sun}$.}
\label{fig:edades_1.5}
\end{figure*}

Results are illustrated in Fig.~\ref{fig:edades_1.5}, in which these evolutionary sequences having 
canonical and thin H envelope are depicted in the $\log(g)$ - $T_{\rm eff}$ plane.
We have also included in the figure the location of a sample of ELM WDs and also the 
known ELMVs 
\citep{2012ApJ...750L..28H,2013ApJ...765..102H,2013MNRAS.436.3573H,2015MNRAS.446L..26K,2015ASPC..493..217B,2016ApJ...818..155B,2017ApJ...835..180B}.
The solid black line to the left
connects points of equal age (13.7 Gyr) for the canonical tracks, while
the solid lines to the right connect points of equal age (13.7 Gyr) but
for the thin H-envelope tracks. The dashed lines also correspond to
the Galactic disk age limit, and are only marked as projections to delimit
the region. The main result illustrated by Fig.~\ref{fig:edades_1.5} is the 
existence of a region named "Thin H envelope" in which only
ELM WDs with thin H envelope can be found. In fact, ELM WDs with
canonical H envelopes are not expected to evolve to $T_{\rm eff}$ values lower than the limit imposed by 
the left solid line. Only ELM WDs with thin
H envelopes, and hence short cooling ages, are expected to evolve to such
lower $T_{\rm eff}$ (but not lower to the right solid line).
More specifically, low-mass WDs with $M_{\star} < 0.20\ M_{\sun}$
with residual H burning (canonical envelopes)  should
be no cooler than roughly $7000\ $K, as observations seem to show. But if such
low-mass WDs were born with thinner H envelope (for instance, due to unstable
mass transfer) they could be expected at much lower effective temperatures
(down to $\sim 2500\ $K). This means that the detection of low-mass WDs in this
region could be indicative that such stars were formed with thin H
envelopes. One possibility is through unstable mass loss via 
common-envelope episodes. In Fig.~\ref{fig:edades_1.5} we 
have also marked  a zone named
"Forbidden region" where ELM WDs are not expected to be found because
to reach such region, evolutionary times larger than the Galactic disk age would
be needed (even assuming ELM WDs with thin H envelopes).

Another observable consequence of a reduction of the H envelope concerns its
impact on the surface gravity.
It is clear from Fig.~\ref{fig:edades_1.5} and also from
Table~\ref{tab:cooling}  that reducing the thickness of the H envelope yields evolutionary sequences with gravity values substantially higher. This is because a reduction in the H content 
cause the envelope to become denser, thus implying a decrease
in the stellar radius.  At variance with mass average
WDs, where the mass of residual H
is much smaller, the impact on the gravity
becomes more significant in low-mass He-core WDs. The increase in
gravity resulting from a reduction of the H envelope has  
consequences for the spectroscopic determination of the star parameters. At first glance, the uncertainty in the stellar mass
due to the uncertainty in the thickness of the H envelope is at least of $\sim 0.025\ M_{\sun}$. 

\begin{figure*}  
\centering  
\includegraphics[clip,width=420pt]{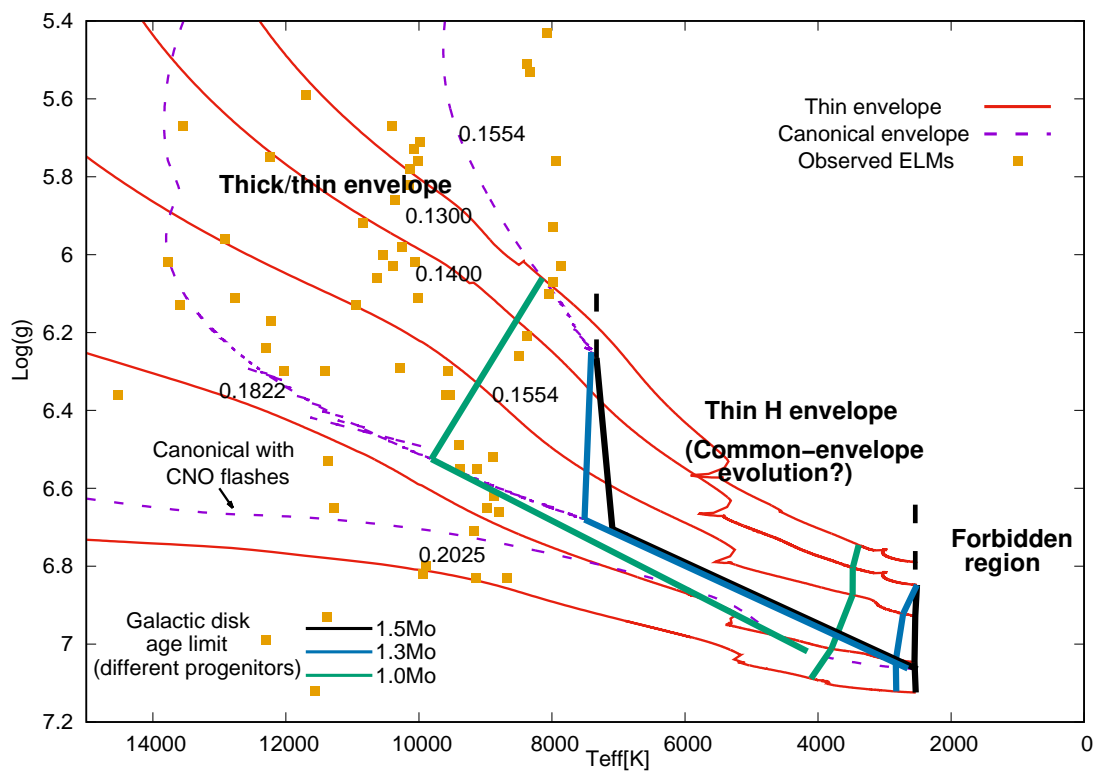}  
\caption{Same as Fig.~\ref{fig:edades_1.5} but for different masses of the
  progenitor star ($1.0$, $1.3$ and $1.5\ M_{\sun}$), marked with different
   solid colored lines.}
\label{fig:edadesdif}
\end{figure*}

 The above mentioned expectations are based on the asssumption that the progenitor star has the maximum value of the initial mass, and hence the shortest ages,  to form ELM WDs, i.e,  $1.5\ M_{\sun}$. Now, we explore how
the conclusions are altered if the mass of the progenitor star is changed.
In particular, results for ELM WDs coming from $1.0$ and $1.3\ M_{\sun}$ progenitor stars are shown in Fig.~\ref{fig:edadesdif}. Note that lines
of equal ages shift now to higher $T_{\rm eff}$ values in comparison with the
lines predicted in the case of the $1.5\ M_{\sun}$ progenitor, as it should be expected because of the larger progenitor ages.
However, this shift is not too significant and
we can conclude that these boundaries are not too sensitive to the mass of the
progenitor. Specifically, if low-mass WDs with $M_{\star} < 0.18\ M_{\sun}$
were formed with thick H envelope and came from a progenitor with
$1.3\ M_{\sun}$, they should not be cooler than $\sim 7400\ $K,
and if such low-mass WDs were born with thin H envelope, they could
reach down to  $\sim 2700\ $K. For the $1.0\ M_{\sun}$ progenitor, the limits
are roughly $8100\ $K and $3500\ $K, for the thick and thin H envelope,
respectively.
The region where only thin H envelope can be found remains mostly unchanged and well defined.

 Finally, note that all of our thin H envelope sequences experience
a sort of hook in their track between $T_{\rm eff} \sim 5000 - 6000\ $K. 
The reason for this is related to the deepening of the convection zone that  reaches the
H/He transition region, enriching the envelope with He and hence, making
the star more compact.

\section{Summary and conclusions}  
\label{conclusions}

In this work, we  have presented new evolutionary sequences of ELM WDs
with thin H envelope and we have studied the differences in the
cooling times comparing thin and thick H-envelope sequences.
ELM WD stars are thought to be the result of
strong  mass-loss episodes in interactive binary systems before the
He flash during the red giant branch phase of low-mass  stars
\citep{2000MNRAS.316...84S,2013A&A...557A..19A,2016A&A...595A..35I}. It
is currently accepted that these stars are formed  either through
unstable mass loss, for instance, via common-envelope episodes,
or stable mass loss
via Roche-lobe  overflow  in close binary systems \citep[see  for   a
  recent  discussion][]{2016A&A...595A..35I}. All of the
existing studies \citep[for
  instance,][]{2013A&A...557A..19A,2016A&A...595A..35I} are performed
considering ELM WDs evolving from progenitors that have experienced
stable mass transfer, leading to thick H envelopes and hence, to very
long cooling ages due to residual H burning. However, we cannot rule out the
scenario where these stars are formed through unstable mass loss,
leading to ELM WDs with thin H envelopes unable to sustain residual H
burning, resulting in much shorter cooling ages. 
%

In the present paper, we determine the lowest $T_{\rm eff}$
that could be reached by ELM WDs. We studied the cooling times of ELM WD
models with canonical (thick) H envelopes taken from \cite{2013A&A...557A..19A},
and also ELM WD models characterized by thin H envelopes. This last set of models was generated on the basis of models with thick envelopes, for which we artificially reduced the 
thickness of the H envelope at high luminosities.
We stress that the reduction in the thickness of the H envelope does not
need to be large for the nuclear burning to become negligible:
for instance, for the sequence with $0.1554\ M_{\sun}$, only a factor
of reduction of $\sim 2.4$ is enough. 
We analyzed the cooling ages of our
WD sequences considering the Galactic disk age, and assumed that stars were born at the same time of the formation of the disk, so they
would have had enough time to reach the minimum $T_{\rm eff}$.  We
also considered the age of some possible progenitors that can form ELM
WDs.
We found that there is a well limited region in the $\log(g)$ vs $T_{\rm eff}$ 
plane where only ELM WD stars with thin H envelope can be
found. We also found that changing the progenitor mass does not
significantly change the lowest $T_{\rm eff}$s reached by these stars.
Additionally, we found that a reduction of the H envelope has an important
observational impact, because it increases the value of $g$, leading
to consequences in the spectroscopic determinations for these stars.

Our results show that it would be possible to find ELM WD stars at
very low $T_{\rm eff}$ if they had a thin H envelope. This being the case, we could conclude that these stars experienced unstable mass loss
during their previous evolution, shedding light on the formation of
such stars. We are aware that the detection of cool ELM WDs is a
difficult task, due to the inaccuracy of the spectroscopic technique
below $7000\ $K. However, hopefully with GAIA parallax measurements,
we might be able to identify such cool ELM WDs.

  
\begin{acknowledgements}
We wish to thank our anonymous referee for the constructive
comments and suggestions that greatly improved the original version of
the paper. We also wish to warmly thank the helpful comments of M. Kilic,
W. R. Brown and K. J. Bell. Part of this work was supported by AGENCIA
through the Programa de Modernizaci\'on Tecnol\'ogica BID 1728/OC-AR,
and by the PIP 112-200801-00940 grant from CONICET. This research made use 
of NASA Astrophysics Data System.
\end{acknowledgements}

\bibliographystyle{aa}
\bibliography{biblio}

\end{document}